\documentclass[%
reprint,
showpacs,
bibnotes,
amsmath,amssymb,
aps,
prb,
floatfix,
]{revtex4-1}

\newlength{\myfigwidth}
\newlength{\myownfigwidth}
\usepackage{natbib}
\usepackage{xcolor}
\usepackage{amsmath}
\usepackage{graphicx}
\usepackage{cancel}
\usepackage{epsfig}
\usepackage{mathrsfs}
\usepackage{amssymb}
\usepackage{relsize}
\usepackage{marginnote}
\usepackage[breaklinks]{hyperref}
\usepackage{hyperref}

\usepackage[shortlabels]{enumitem}
\usepackage{tabularx}

\newcommand{\bcen}{\begin{center}}
\newcommand{\ecen}{\end{center}}
\newcommand{\btab}{\begin{tabular}}
\newcommand{\etab}{\end{tabular}}
\newcommand{\bdes}{\begin{description}}
\newcommand{\edes}{\end{description}}

\newcommand{\beq}{\begin{equation}}
\newcommand{\eeq}{\end{equation}}
\newcommand{\bea}{\begin{eqnarray}}
\newcommand{\eea}{\end{eqnarray}}
\newcommand{\non}{\nonumber}

\newcommand{\bary}{\begin{array}}
\newcommand{\eary}{\end{array}}
\newcommand{\benum}{\begin{enumerate}}
\newcommand{\eenum}{\end{enumerate}}
\newcommand{\bitem}{\begin{itemize}}
\newcommand{\eitem}{\end{itemize}}
\newcommand{\bn} { \mbox{\boldmath $n$}}

\newcommand{\bp} { \mbox{\boldmath $p$}}
\newcommand{\br} { \boldsymbol{r} }

\newcommand{\bx} { \boldsymbol{x}}

\newcommand{\bz} { \mbox{\boldmath $z$}}

\newcommand{\bL} { \mbox{\boldmath $L$}}

\newcommand{\bV} { \mbox{\boldmath $V$}}

\newcommand{\eqn}[1] {Eq.~(\ref{#1})}

\newcommand{\Sect}[1] {Sec.~\ref{#1}}
\newcommand{\fig}[1]{Fig.~\ref{#1}}
\newcommand{\Fig}[1]{Fig.~\ref{#1}}
\newcommand{\myonlinecite}[1] {[\onlinecite{#1}]}

\newcommand{\sun}{{\hbox{$\odot$}}}
\newcommand{\dummyfigure}[1]{{}}

\newcommand{\MNRAS}{Mon. Not. R. Astron. Soc.~}

\newcommand{\mylabel}[1]{\label{#1}}

\bibpunct[, ]{}{}{,}{s}{,}{,}

\begin{document}
\baselineskip=17.2pt

\relax

\title{A geometric method to locate Neptune}
\author{Siddharth Bhatnagar}
\email{siddharth.bhatnagar16ug@apu.edu.in}
\author{Jayanth Vyasanakere P.}
\email{jayanth.vyasanakere@apu.edu.in}

\affiliation{The School of Arts and Sciences, Azim Premji University, Bengaluru, 562 125, India}

\author{Jayant Murthy}
\email{jmurthy@yahoo.com}
\affiliation{Indian Institute of Astrophysics, Bengaluru, 560 034, India}

\date{\today}

\begin{abstract}
We develop a direct geometric method to determine the orbital parameters and mass of a planet, and we then apply the method to Neptune using high-precision data for the other planets in the solar system. The method is direct in the sense that it does not involve curve fitting. This paper, thereby, offers a new pedagogical approach to orbital mechanics that could be valuable in a physics classroom.
\end{abstract}

\maketitle


\section{Introduction}
\mylabel{sec:Intro}
The discovery of Neptune in 1846 by observing deviations of Uranus from its predicted orbit remains one of the crowning achievements of Newtonian mechanics.\cite{CE,LeVerrier1846,Adams1846} The discoverers faced a difficult inverse problem, easily solvable today with electronic computers, but Herculean in the mid-nineteenth century.\cite{Rines1912,Smart1946} The discoverers, Urbain Le Verrier and John Couch Adams, independently investigated Alexis Bouvard's ephemeris tables for Uranus to check their accuracy and found no computational errors.\cite{Smart1946, Rines1912,Bhatnagar2018} With this, it became clear that the problem lay not with the tables, but rather with the planet itself! The mathematical problem of locating the precise position and determining the mass of the perturbing object was non-trivial; \cite{Smart1946, Bhatnagar2018} Le Verrier and Adams proceeded by linearizing the system of equations involved and solving them by least squares. Computing the semi-major axis of the perturbing body was also a complex problem, for which an informed estimate had to be made. This was done using the Titius-Bode law,\cite{Nieto2014} which held for the five planets known since antiquity. Neptune has subsequently been ``rediscovered'' with simpler methods several times after its real discovery.\cite{Brown1931,Lyttleton1960,Brookes1972,Lai1990,Eriksson2018,Bhatnagar2018}

In today's age of high-precision data, orbital state vector data of most solar system bodies are available.\cite{Horizon} We have retrieved these data for the major planets in {\em Cartesian coordinates}, with the origin at the instantaneous center of the Sun (see the Appendix for more details). These data, collected at a time step of 2~h, range from the date of Uranus' discovery (March 1781) to the present day (March 2020). It is clear that we have an advantage over the discoverers, in terms of the nature, quality and duration of data at hand. Using this advantage, we describe a purely geometric way of locating a planet, focusing on Neptune.

Since we work with state vector data and not the orbital parameters of the known planets, our approach naturally lends itself to vector analysis. The geometrical ideas are encoded into scalar (dot) products and vector (cross) products. The method emphasizes basic ideas of classical mechanics, such as Newton's laws, Kepler's laws, and motion in the presence of an inverse square law force, but does not use curve fitting to locate the unknown planet. All these ideas are typically encountered in an undergraduate physics curriculum, making our method better suited to pedagogical purposes than the historical ones.

\section{Equation of Motion}
Our model comprises \textit{only} nine objects: the Sun, the seven known planets at the time (Mercury to Uranus) and Neptune, which we consider unknown for the purpose of demonstrating our method.\cite{Asteroid} Newton's law written for the $i$th object reads
\beq
\frac{d^2 \bx_i}{dt^2}=-\sum_{j \neq i} G M_j \frac{\bx_i-\bx_j}{|\bx_i-\bx_j|^3} \qquad (i=1 \text{ to }9),
\mylabel{eqn:Newton}
\eeq
where $G$ is the universal gravitational constant, $\bx_i$~s and $\bx_j$~s stand for the position of the center of mass (COM) of the objects (including their moons, if any), and $M_j$~s are the masses (again including the moons, if any). However, we need not worry about solving these coupled differential equations, for we have the solution in terms of data for all objects except Neptune, whose parameters need to be determined using these equations.

The $d^2 \bx_i/dt^2$ in \eqn{eqn:Newton} denote accelerations with respect to an inertial frame. The COM of the solar system is a reasonably good choice for such a frame. However, the COM is not an observable, but rather a construct that can only be obtained after having knowledge of \textit{all} constituents of the system, including Neptune ($N$), which is yet to be ``discovered.'' We circumvent this problem by subtracting \eqn{eqn:Newton} written for the Sun ($\sun$) from the same equation written for Uranus ($U$). This gives
\beq
G M_N \left( \frac{\br_N - \br_U}{|\br_N-\br_U|^3} - \frac{\br_N}{|\br_N|^3} \right) = \bV(t),
\mylabel{eqn:Voft}
\eeq
where 
\bea
\bV(t) &=& \frac{d^2 \br_U}{d t^2} + G \left( M_\sun + M_U \right) \frac{\br_U}{|\br_U|^3} \non \\
&& - \sum_{j \neq \sun, U, N} G M_j \left( \frac{\br_j-\br_U}{|\br_j-\br_U|^3} - \frac{\br_j}{|\br_j|^3} \right) 
\mylabel{eqn:Voft_Expansion}
\eea
and where $\br_i=\bx_i-\bx_\sun$, which are relative coordinates with respect to the Sun. It is instructive to write down all the terms explicitly to understand the derivation of Eqs.~\ref{eqn:Voft} and \ref{eqn:Voft_Expansion}, and to note that $\br_i-\br_j=\bx_i-\bx_j$.

In \eqn{eqn:Voft_Expansion}, the $\br_i$~s are observable without \textit{a priori} knowledge of Neptune. Thus, with \eqn{eqn:Voft} and the assembled data from Ref.~\onlinecite{Horizon} (as elaborated in \Sect{sec:Intro}), $\bV(t)$ is known; see \Sect{sec:UV} for a physical understanding of this vector. The stage is now set to find the unknowns $\br_N(t)$ and $M_N$, which we do through a purely geometric approach.

\section{Understanding the vector $\bV$}
\label{sec:UV}

Since our discussion revolves around the vector $\bV$, it is important to discuss what it means physically. From \eqn{eqn:Voft}, $\bV(t)$ would be zero if Neptune did not exist; it is that part of the acceleration of Uranus not explained by the other planets. 

Consider $\hat{\bV}$, which is the unit vector along $\bV$. In \fig{fig:Vcap}, this is denoted by a red solid vector. If $\br_U$ is imagined to be held fixed and $\br_N$ is made to rotate anticlockwise once around the $z$ axis, then, simultaneously, $\hat{\bV}$ will complete two anticlockwise circuits around the $z$ axis. We will revisit $\hat{\bV}$ in \Sect{sec:ConjOpposPrecise}.

\begin{figure}[]
		\centerline{\includegraphics[height=4.0cm,width=4.0cm,trim=48 36 34 64,clip]{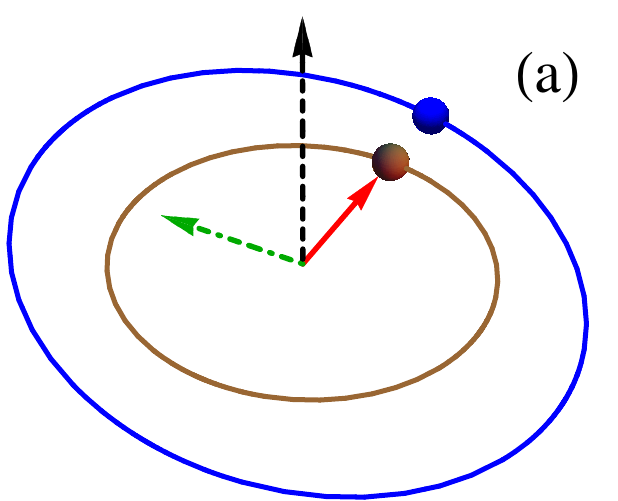} ~~~~ \includegraphics[height=4cm,width=4.0cm,trim=48 36 34 64,clip]{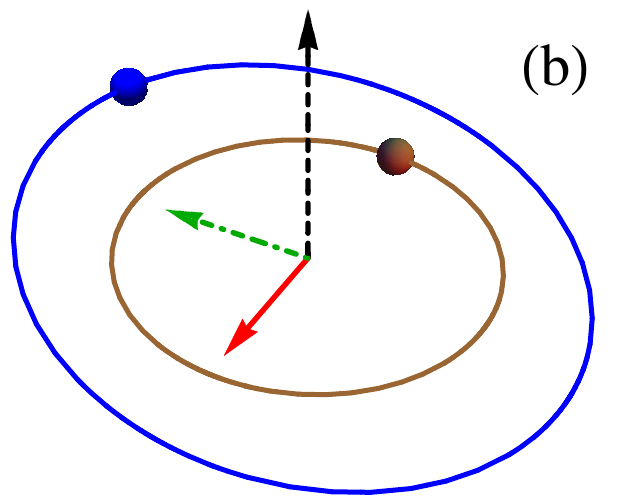}}
		\centerline{\includegraphics[height=4.0cm,width=4.0cm,trim=48 36 34 64,clip]{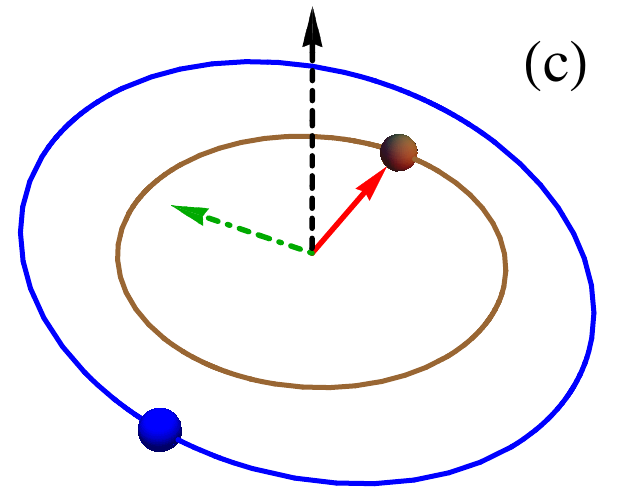} ~~~~ \includegraphics[height=4cm,width=4.0cm,trim=48 36 34 64,clip]{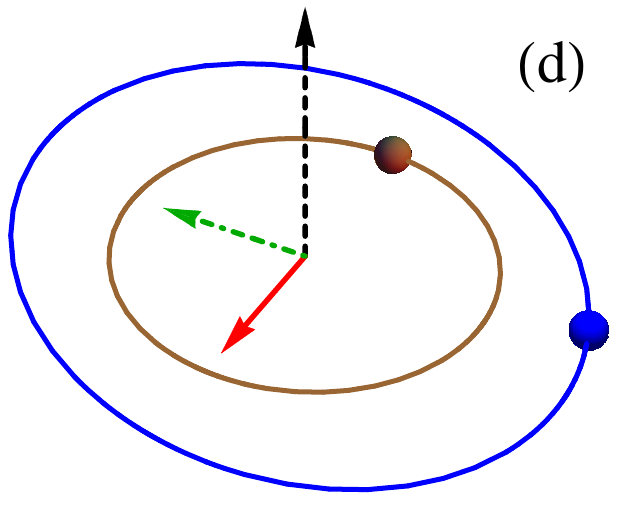}}
	\caption{The behavior of $\hat{\bV}$ as Neptune (blue sphere---the one on the outer orbit) moves around in its orbit, with Uranus (brown sphere---the one on the inner orbit) held fixed. The Sun (not shown) is at the center and the black dashed line is the $z$-axis. The orbits of Uranus (brown inner orbit) and Neptune (blue outer orbit) are both shown to be slightly inclined to the $x$--$y$ plane. The red solid line denotes $\hat{\bV}$. The green dot-dashed line denotes vector $\bp$ (\eqn{eqn:vecp}). (a) and (c)  correspond to conjunction and opposition of the Sun--Uranus--Neptune system, respectively. (b) and (d) correspond to configurations, which are not conjunction or opposition, but still satisfy $\chi=0$ (see \eqn{eqn:chi} and the discussion around it). Note that this figure is a schematic for the sake of visualizing $\hat{\bV}$ and was not derived from any data.}
	\mylabel{fig:Vcap}
\end{figure}

We now turn our attention to $V(t)=|\bV(t)|$, the magnitude of $\bV$. This is shown in \Fig{fig:NV}.\cite{Noise} With $\bV$ being that part of Uranus' acceleration arising exclusively from Neptune, the peaks in \fig{fig:NV} are expected to correspond to conjunctions. This can be confirmed by modeling the orbits of Uranus and Neptune as co-planar circles around the Sun, and obtaining from \eqn{eqn:Voft} that $V$ peaks during conjunction, irrespective of the radii of the orbits. Hence, it is reasonable to expect that the peaks in \fig{fig:NV}, which occurred in November 1822 and June 1994, signify conjunctions. However, this prediction is not precise as the planets' orbits are neither exact circles nor exactly co-planar. Therefore, the analysis, thus far, provides but a rough estimate for the date of conjunction.

\begin{figure}[]
	\centerline{\includegraphics[height=5cm]{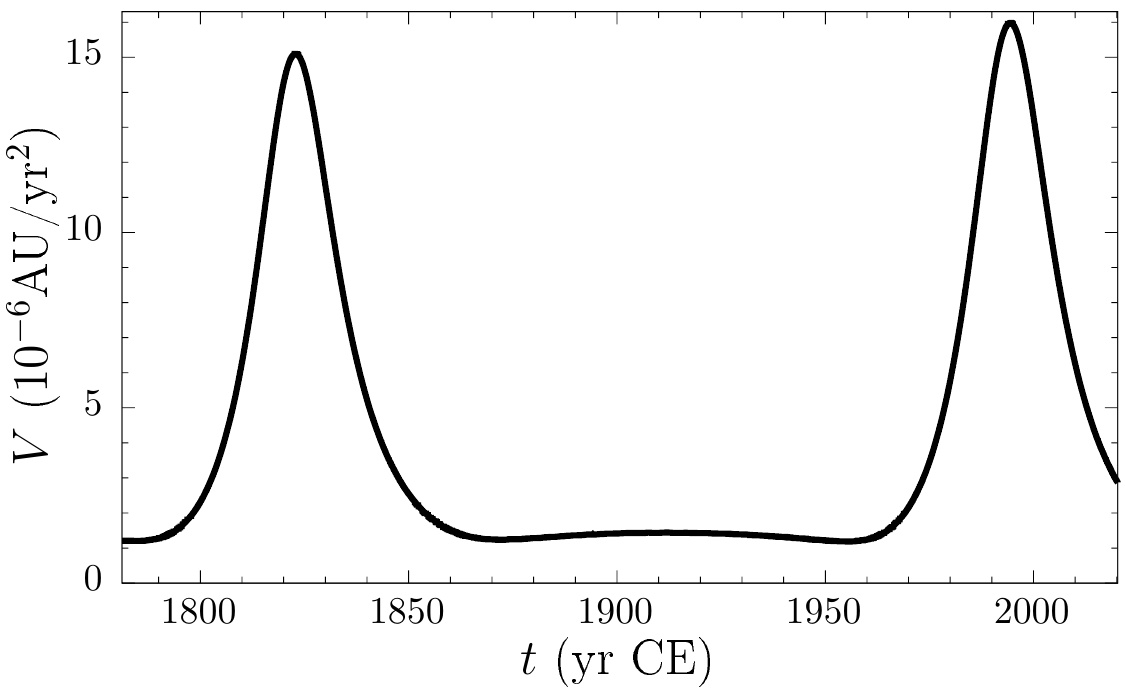}}
	\caption{$V$, which is the magnitude of $\bV$ (\eqn{eqn:Voft_Expansion}), is shown as a function of time. The two peaks correspond to November 1822 and June 1994. As argued in \Sect{sec:UV}, these peaks provide a rough estimation of the time of conjunction of the Sun--Uranus--Neptune system.}
	\mylabel{fig:NV}
\end{figure}

\section{Determination of Conjunction and Opposition with Uranus}
\label{sec:ConjOpposPrecise}

\begin{figure}[]
	\centerline{\includegraphics[height=5cm]{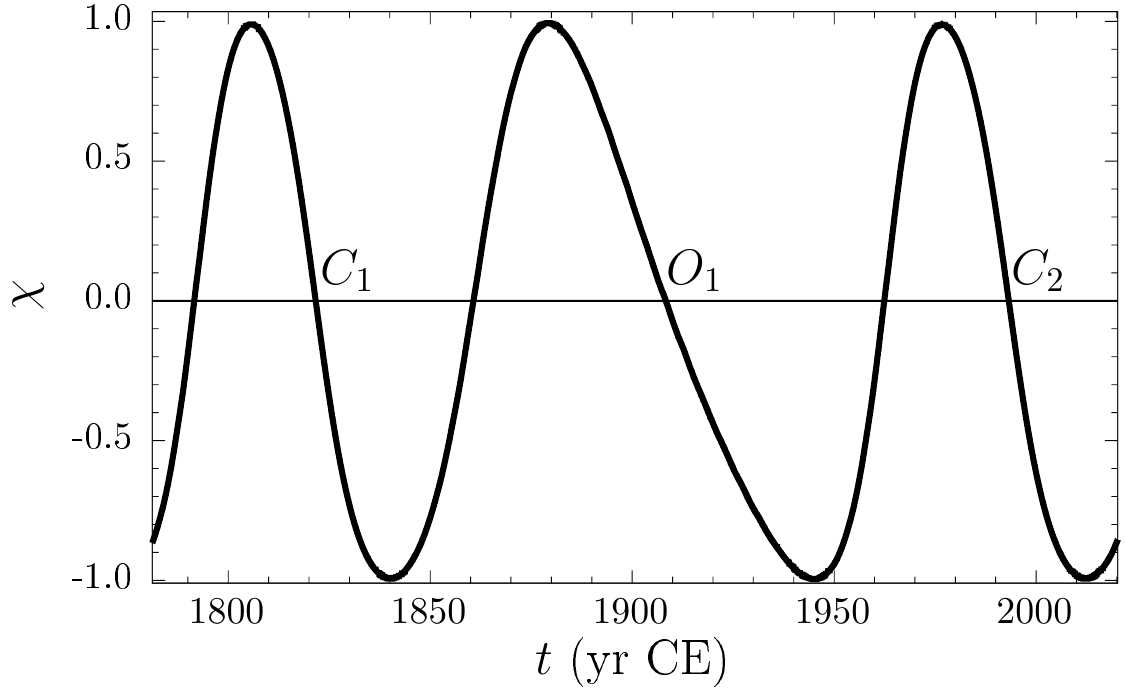}}
	\caption{Precise determination of the conjunction and opposition of the Sun--Uranus--Neptune system. $\chi$ (\eqn{eqn:chi}) is shown as a function of time. Among the several roots of $\chi$ in the duration shown, the ones occurring in August 1821 and March 1993 determine conjunctions ($C_1$ and $C_2$), while the one in May 1908 determines opposition ($O_1$). The origin of the remaining roots (also depicted in parts (b) and (d) of \fig{fig:Vcap}) is discussed below \eqn{eqn:chi}.}
	\mylabel{fig:chi}
\end{figure}

To obtain precise conjunction and opposition times, we recall that in a spherical polar coordinate system with the Sun at the origin, the azimuthal angle ($\phi$) for both Uranus and Neptune will be the same during conjunction, and differ by $\pi$ radians at opposition.
From \eqn{eqn:Voft} and \fig{fig:Vcap}, it is clear that during {\em both} conjunctions {\em and} oppositions, the vector $\bV$ will have the same $\phi$ coordinate as Uranus. This observation motivates us to define a vector 
\beq
\bp(t) = \hat{\bz} \times \hat{\br}_U(t), 
\label{eqn:vecp}
\eeq
which lies in the $x$--$y$ plane and is normal to $\br_U(t)$ (denoted as a green dot-dashed line in \fig{fig:Vcap}). In terms of $\bp$, we can hope to determine a conjunction or an opposition by checking whether $\bV$ is perpendicular to $\bp$ at a given time.
In other words, we define a quantity $\chi$, which is a measure of the projection of $\hat{\bV}$ along $\bp$, as
\beq
\chi(t) = \hat{\bV}(t) \cdot \bp(t),
\mylabel{eqn:chi}
\eeq
and look for its roots. 

Figure \ref{fig:chi} shows $\chi$ as a function of time. As expected, $\chi$ does vanish around the estimated times for conjunctions and oppositions as obtained in the \Sect{sec:UV} by looking at the peaks in \Fig{fig:NV}. We have already seen that the $\phi$ coordinate of $\bV$ is the same as that of Uranus during both conjunctions and oppositions. This observation can also be used to further verify whether a given root of $\chi$ corresponds to either a conjunction or an opposition.

The remaining roots of $\chi$, which do not correspond to conjunctions or oppositions can be understood as follows. During conjunctions and oppositions, the projections of both the bracketed terms in \eqn{eqn:Voft} on $\bp$ vanish. However, $\chi$ can also vanish if the projections of these two terms cancel each other. The relative dominance of these terms flips when the system moves from a conjunction to an opposition or vice versa, resulting in an extra root between them. Basically, if the $n$th root corresponds to a conjunction, the $(n+2)$th root will be an opposition, the $(n+4)$th root will again be a conjunction, and so on. 

In this manner, we conclude that the roots of $\chi$ in \Fig{fig:chi}, which happen to fall in August 1821 and March 1993, are conjunctions and designate them as $C_1$ and $C_2$, respectively. The root falling in May 1908 then corresponds to an opposition, which we call $O_1$.

\section{Determination of the orbital period and semi-major axis}
\label{sec:OrbTimePeriod}

The synodic period of Neptune with respect to Uranus is the time difference between the conjunctions $C_1$ and $C_2$: $T_{sy} = 171.6$ years. If Uranus and Neptune were indeed co-planar and moved at constant angular speeds, this information, along with the orbital period of Uranus, would be sufficient to calculate the sidereal period of Neptune around the Sun to be $T_N \approx 165$ years. However, this is an approximation.

\begin{figure}[]
	\centerline{\includegraphics[height=5cm]{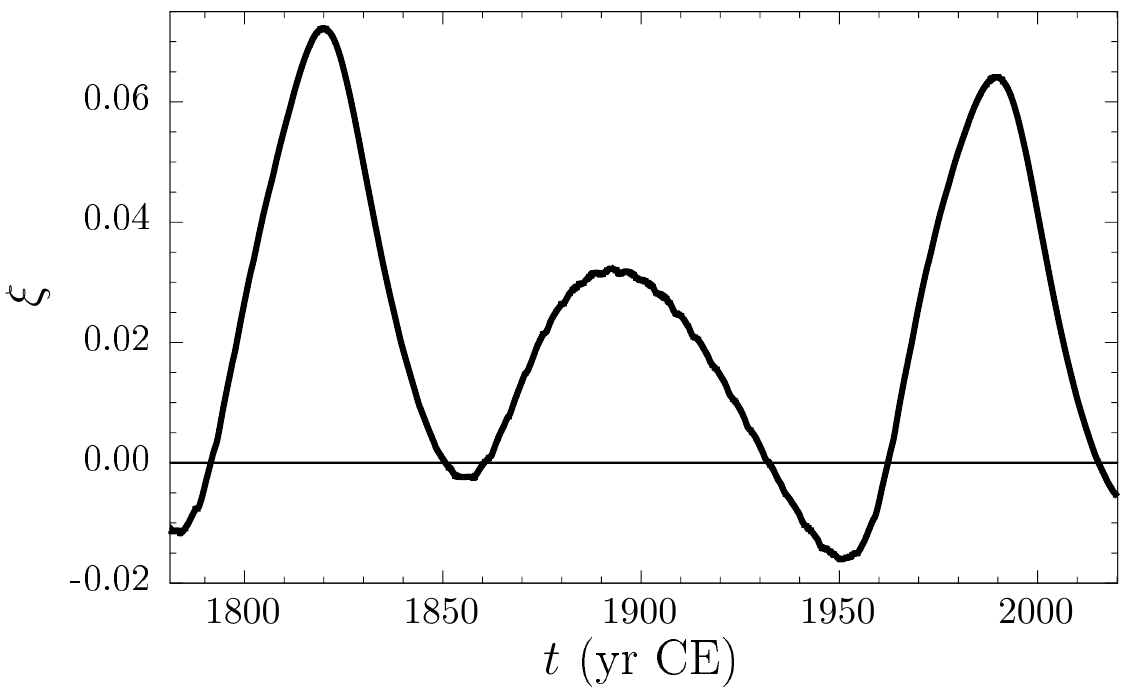}}
	\caption{The projection of $\bV(t)$ along the normal to the orbital plane of Uranus, quantified through $\xi$ (\eqn{eqn:xi}), is shown as a function of time. The roots of $\xi$ in the duration shown occur at June 1791, September 1850, June 1860, March 1932, April 1962, and May 2015. As argued in \Sect{sec:OrbTimePeriod}, the roots in September 1850 and May 2015 can be used to precisely determine the orbital period of Neptune.}
	\mylabel{fig:xi}
\end{figure}

We, therefore, venture to obtain $T_N$ precisely by defining
\beq
\xi(t) = \hat{\bV}(t) \cdot \hat{\bn}_U,
\mylabel{eqn:xi}
\eeq
where $\hat{\bn}_U$ is a unit vector along $\left(\br_U \times \frac{d \br_U}{dt} \right)$ evaluated at our model's epoch, March 1781 (referred to hereafter as $T_I$). In other words, $\hat{\bn}_U$ is a unit normal to the orbital plane of Uranus. The quantity $\xi$, a measure of projection of $\bV(t)$ along $\hat{\bn}_U$, is shown in \Fig{fig:xi} as a function of time. When Neptune crosses the orbital plane of Uranus, $\xi$ vanishes. However, similar to the roots of $\chi$, between any two consecutive roots of this kind will exist another kind of root, where the components along $\hat{\bn}_U$ of the two terms bracketed in \eqn{eqn:Voft} cancel each other. Thus, with reference to \fig{fig:xi}, if the time difference between the $n$th and $(n+4)$th roots of $\xi$ corresponds to $T_N$, then that between the $(n+1)$th and $(n+5)$th roots corresponds to $T_{sy}$, and vice versa. 
In particular, the roots in June 1791 and April 1962 differ by 170.8 years, which is very close to $T_{sy}$. Hence, the roots in September 1850 and May 2015 can be used to give $T_N=164.7$ years. The semi-major axis ($a$) is then obtained from $T_N$ using Kepler's third law (assuming $M_N \ll M_\sun$) to be 30.05 astronomical units (AU). 

\section{Determination of other orbital parameters and mass}
\label{sec:OtherOrbParam}

To begin here, the polar angle $\theta$ and the azimuthal angle $\phi$ (the spherical polar coordinates) of Neptune during a conjunction with Uranus can be determined as follows.

Equation \ref{eqn:Voft} can be re-expressed as

\vspace{-0.4cm}
{\small
\beq
\br_N = \br_U + |\br_N-\br_U|^3 \left(\frac{\br_N}{|\br_N|^3} + \left| \frac{\br_N - \br_U}{|\br_N-\br_U|^3} - \frac{\br_N}{|\br_N|^3} \right| \hat{\bV} \right).
\mylabel{eqN:ItnConj}
\eeq
}
As an initial guess, we can take $\theta$ for Neptune to be $\pi/2$ radians, and, since we are dealing with a conjunction, $\phi$ of Neptune can be taken equal to that of Uranus. By freezing Neptune's $r$ coordinate to its semi-major axis $a$, which has already been obtained, we have a guess for $\br_N$. With this, every quantity on the RHS of \eqn{eqN:ItnConj} is known. Hence, the LHS of this equation can be taken as a better estimate for $\br_N$. This way, we can iterate \eqn{eqN:ItnConj} to obtain $\phi$ and $\theta$ of Neptune at conjunction.

A similar procedure can be carried out for an opposition by re-expressing \eqn{eqn:Voft} as 

\vspace{-0.4cm}
{\small
\beq
\br_N = |\br_N|^3 \left( \frac{\br_N - \br_U}{|\br_N-\br_U|^3} - \left| \frac{\br_N - \br_U}{|\br_N-\br_U|^3} - \frac{\br_N}{|\br_N|^3} \right| \hat{\bV} \right)
\mylabel{eqN:ItnOpps}
\eeq
}
and taking  $\phi$ for  Neptune to be $\pi$ radians away from that of Uranus. 

Note that we have re-expressed \eqn{eqn:Voft} in two different ways: one for conjunction (\eqn{eqN:ItnConj}) and the other for opposition (\eqn{eqN:ItnOpps}). These choices ensure convergence of the iteration algorithm in the respective cases. The converged values of $\phi$ and $\theta$ of Neptune after iteration are given in Table~\ref{tab:Direction_Info}. Thus, we obtain the directions of Neptune during $C_1$, $O_1$, and $C_2$.

\begin{table}[]
	\caption{The converged directions (as elaborated in \Sect{sec:OtherOrbParam}) of conjunctions (Uranus and Neptune aligned on the same side of the Sun) and opposition (Uranus and Neptune aligned on the opposite sides of the Sun). $\phi$ and $\theta$ are the spherical polar coordinates, while $\psi$ is the angle in the orbital plane from $C_1$.}
	{\footnotesize
		\centerline{
			\renewcommand{\arraystretch}{1.5}
			\begin{tabular}{c c c c c}
				\hline \hline
				& ~~ Date (year CE) ~~ & ~~ $\phi$ (rad) ~~ & ~~ $\theta$ (rad) ~~ & ~~ $\psi$ (rad) ~~ \\
				\hline
				$C_1$ & August 1821 & 4.532 & 1.512 & 0 \\
				$O_1$ & May 1908 & 1.582 & 1.647 & 3.334 \\
				$C_2$ & March 1993 & 4.787 & 1.490 & 0.256 \\
				\hline \hline
			\end{tabular}
		}
	}
	\mylabel{tab:Direction_Info}
\end{table}

At this stage, we assume that the orbit of Neptune is an ellipse with the Sun at one focus (Kepler's first law). Since the time difference between $C_1$ and $C_2$ is greater than the orbital period, it is clear that $C_2$ is ahead of $C_1$ in the orbit. The orbital plane is characterized by the inclination ($i$) of its normal from the $z$ axis and the longitude of the ascending node ($\Omega$). These can be evaluated from the knowledge of the directions of $C_1$ and $C_2$ as $i=0.1124$ and $\Omega=3.9806$, both in radians. The direction of $O_1$ should be very close to this plane. A negligible component perpendicular to this plane, if any, can be dealt with by projecting $O_1$ onto the plane.
Let $\psi$ be the angle measured in the plane of the ellipse from $C_1$. $\psi$ for $C_2$ and $O_1$ can be easily found, since we already know the corresponding $\phi$ and $\theta$. These values of $\psi$ are tabulated in Table~\ref{tab:Direction_Info}.

Using this information, the equation of the ellipse can be written\cite{Goldstein2007} with unknowns being its eccentricity ($\epsilon$) and the distance of Neptune from the Sun ($r_I$) at $T_I$. These unknowns can be obtained by demanding that Neptune's orbit assumes the appropriate $\psi$ values given in Table~\ref{tab:Direction_Info} at the respective times.\cite{Zeta}

\begin{figure}[]
	\centerline{\includegraphics[height=5cm]{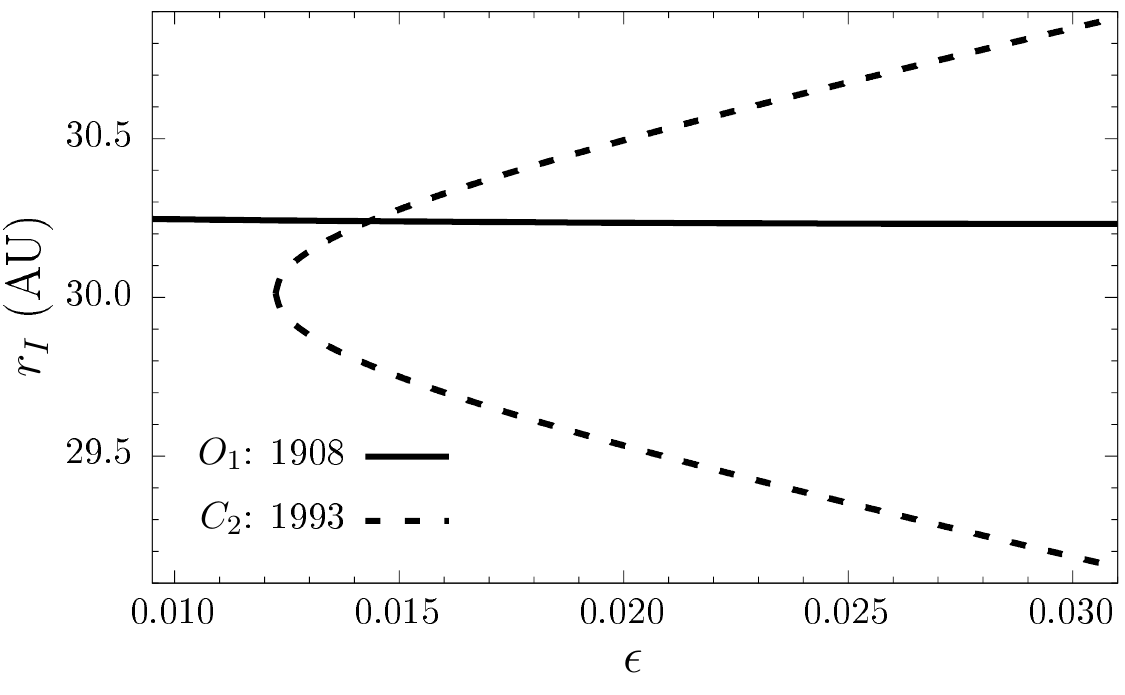}}
	\caption{Determination of eccentricity ($\epsilon$) and the distance of Neptune from the Sun ($r_I$) at our model's epoch ($T_I$). The solid line represents the combination of $\epsilon$ and $r_I$ that is consistent with $\psi$ during $O_1$. The dashed line is the combination of $\epsilon$ and $r_I$ that is consistent with $\psi$ during $C_2$. Their intersection determines $\epsilon$ and $r_I$ of Neptune's orbit.}
	\mylabel{fig:r0vsEcc}
\end{figure}

The combination of $\epsilon$ and $r_I$ that are separately consistent with $\psi$ during $O_1$ and $C_2$ is obtained as shown in \fig{fig:r0vsEcc}. The intersection of these curves gives the required combination of $\epsilon$ and $r_I$. This gives $\epsilon = 0.0143$ and $r_I=30.24$ AU. The orbit of Neptune is thus completely determined. 

The time at which Neptune passes through the perihelion ($T_p$) and the argument of the perihelion (angular coordinate of the perihelion in the orbital plane from the ascending node, denoted by $\omega$) emerge as November 1892 and 3.282 radians, respectively.

Thus, we now know {\em all} the orbital parameters of Neptune, from which its location at any time can be determined. 

The only remaining parameter is the mass of Neptune. From \eqn{eqn:Voft}, this can be written as 
\beq
G M_N = V/\left| \frac{\br_N - \br_U}{|\br_N-\br_U|^3} - \frac{\br_N}{|\br_N|^3} \right|.
\mylabel{eqn:massItn}
\eeq
Now that we know $\br_N$, we can substitute it into \eqn{eqn:massItn} to obtain $M_N$. The mass should of course be a constant, but any deviation of the position obtained from the exact value will cause a variation in $M_N$. Hence, we averaged the $M_N$ values so obtained over the orbital period to arrive at $M_N=1.033 \times 10^{26}$ kg.

\section{Comparison with Actual Values}
\label{sec:Comparison}

\begin{table*}[]
	\caption{Comparison of Neptune's orbital parameters and mass, obtained using the method illustrated in this paper, with the actual values of the osculating orbit (as elaborated in \Sect{sec:Comparison}) in the time interval considered.}
	{\footnotesize
		\centerline{
			\renewcommand{\arraystretch}{1.5}
			\begin{tabular}{l c c c c c c c}
				\hline \hline
				& $i$ (rad) & $\Omega$ (rad) & $a$ (AU) & $\epsilon$ & $\omega$ (rad) & $T_p$ (year CE) & $M_N$ (kg) \\
				\hline
				Our method & 0.1124 & 3.9806 & 30.05 & 0.0143 & 3.282 &  Nov 1892 & $1.033 \times 10^{26}$ \\
				Actual & ~ 0.1122--0.1123 ~ & ~ 3.9769--3.9778 ~ & ~ 29.93--30.31 ~ & ~ 0.002--0.016 ~ & ~ 1.918--3.633 ~ & ~ Aug 1855--Aug 1901 ~ & $1.025 \times 10^{26}$ \\
				\hline \hline
			\end{tabular}
		}
	}
	\mylabel{tab:Summary}
\end{table*}

If $\br_N(t)$ is the trajectory obtained here and $\br_{N0}(t)$ is the actual trajectory of Neptune given by Ref.~\myonlinecite{Horizon}, one way to characterize the error is
\beq
\text{\% deviation}=\frac{|\br_N(t)-\br_{N0}(t)|}{|\br_{N0}(t)|} \times 100.
\label{eqn:PercentageDeviation}
\eeq
The deviation of $\br_N(t)$ from $\br_{N0}(t)$ varies with time and is always within 1.7\%. As viewed from Earth, the predicted direction of Neptune always lies within one degree of its actual direction.

Another approach to testing the quality of our solution is by characterizing the elliptical orbit through its six elements. The inclination ($i$) of the orbital axis and the longitude of the ascending node ($\Omega$) specify the plane of the orbit. The semi-major axis ($a$) and eccentricity ($\epsilon$) specify the size and shape of the ellipse. The argument of the perihelion ($\omega$) specifies the orientation of the ellipse in its plane. The time at which Neptune passes through the perihelion ($T_p$) specifies the epoch.

At any given time, the instantaneous position ($\br_{N0}(t)$) of Neptune as seen from the Sun is already obtained from Ref.~\myonlinecite{Horizon}. From this, the angular momentum ($\bL$) and the energy ($E$) of Neptune can be calculated. While $\epsilon$ and $a$ can be obtained from $E$ and $|\bL|$, $i$ and $\Omega$ can be obtained from $\hat{\bL}$. The calculation of $T_p$ and $\omega$ is more involved, and can be determined from the eccentric anomaly.\cite{Goldstein2007} Hence, at each instant, the orbital elements can be calculated. If Neptune and the Sun were isolated, the orbital elements would be constants. However, due to perturbations by other planets, they do vary. We summarize these variations in Table \ref{tab:Summary}, which also lists the orbital elements calculated using $\br_N(t)$ obtained from our method. It is apparent that they compare well with the corresponding actual values. Note that since $\br_N(t)$ is presumed to be perfectly elliptical, we obtain definite values for the orbital parameters, not a range of values.

\section{Conclusion}

The discovery of Neptune was a magnificent example of mathematical and scientific analysis. While the older methods were certainly able to locate Neptune, they were less transparent in terms of the usage of physics principles and geometrical arguments than that advanced here.

Aided with modern-age data, the method we have described to locate Neptune is simple, both conceptually and by means of calculation. Moreover, it is also a {\em direct} geometric method, without any curve fitting or solving of differential equations. All steps in this method are based on ideas encountered in an undergraduate curriculum, such as vector analysis and the laws of planetary motion. Thus, our method offers a pedagogical alternative to traditional methods. 

This method can also be applied to the case of a hitherto undiscovered planet. For this, sound data for a long enough duration (at least until one time period of the planet) of {\em all} the other planets in the system are required. This method can also serve to provide an initial guess for more sophisticated analyses, be it for application within our solar system or to exoplanetary systems.

\begin{acknowledgments}
	
The authors thank Rajaram Nityananda of Azim Premji University for illuminating discussions and the three
anonymous reviewers for their helpful suggestions.
	
\end{acknowledgments}

\appendix*

\section{DOWNLOADING THE DESIRED DATA}
The following settings were made on Ref.~\onlinecite{Horizon} to generate and download the data.
\begin{enumerate}[(1)]
\item Choose ``Vectors'' from the ``Ephemeris Type'' menu. This will ``generate a Cartesian state vector table of any (solar system) object with respect to any major body."
\item Choose the object in question from the ``Target Body'' menu.
\item Write ``@sun'' in the box provided in the ``Coordinate Origin'' menu. This will ensure that the coordinate origin is fixed to the center of the Sun at all times.
\item From the ``Time Span'' menu, choose a time span in the appropriate format (specified in the same webpage next to these fields) and the necessary time step.
\item Under the ``Table Settings'' menu, select vector table output as ``Type 2 (state vector \{x,y,z,vx,vy,vz\}).'' Then, choose the required output units from the ``output units'' submenu. Now, choose ``body mean equator and node of date'' from the ``reference plane'' submenu. This will set the reference $x$--$y$ plane to coincide with the Sun's mean equator. Then, choose ``ICRF/J2000.0'' from the ``reference system'' submenu. Leave ``aberrations'' as ``Geometric states (no aberrations; instantaneous ephemeris states).'' Check the ``labels,'' ``CSV format,'' and ``object page'' options.
\item From the ``Display/Output'' menu, check the ``download/save'' option so that the ephemeris results can be saved to a local file.
\end{enumerate}
Then select the ``Generate Ephemeris'' option and a .txt file with the ephemeris data will be downloaded.


\begin{thebibliography}{99}
\bibitem{CE} All dates mentioned in this paper are in CE.
\bibitem{Asteroid} Inclusion of more objects such as big asteroids like Ceres will nominally increase accuracy.
\bibitem{Noise} The short term noise in Figs.~ \ref{fig:NV}--\ref{fig:xi} is removed by taking the moving average.
\bibitem{Zeta} Actually, $\zeta=$ sign of the radial component of velocity at $T_I$ is another unknown parameter while solving for the ellipse. With other parameters fixed, we tried both possibilities ($\zeta = \pm 1$) and have kept $\zeta=+1$ as there is no root analogous to that found in \fig{fig:r0vsEcc} for $\zeta=-1$.
\end{thebibliography}
\end{document}